\documentclass[aps, prl, 10pt, twocolumn, superscriptaddress,noshowpacs, preprintnumbers, longbibliography, bibnotes,hyperref,floatfix]{revtex4-1}

\usepackage[colorlinks=true,breaklinks=true]{hyperref}
\hypersetup{allcolors=[rgb]{0.0 0.0 1.0},linkcolor=[rgb]{0.75 0.05 0.05}}
\usepackage[dvipsnames]{xcolor}
\usepackage{mathtools}
\usepackage{amsmath}
\usepackage{amsfonts}
\usepackage{amssymb}
\usepackage{bbold}
\usepackage{multirow}
\usepackage{bm}
\usepackage{hyperref}
\long\def\exclude#1{}
\usepackage{orcidlink}

\newcommand{\bp}{{\bf p}}
\newcommand{\br}{{\bf r}}
\newcommand{\bk}{{\bf k}}

\newcommand{\bv}{{\bf v}}

\newcommand{\bn}{{\bf n}}

\begin{document}

\title{
Self-Interacting Dark Sectors in Supernovae Can Behave as a Relativistic Fluid}
\author{Damiano F.\ G.\ Fiorillo \orcidlink{0000-0003-4927-9850}} 
\affiliation{Niels Bohr International Academy, NBI,
University of Copenhagen, 2100 Copenhagen, Denmark}

\author{Edoardo Vitagliano
\orcidlink{0000-0001-7847-1281}}
\affiliation{Dipartimento di Fisica e Astronomia, Università degli Studi di Padova,
Via Marzolo 8, 35131 Padova, Italy}
\affiliation{Istituto Nazionale di Fisica Nucleare (INFN), Sezione di Padova,
Via Marzolo 8, 35131 Padova, Italy}

\begin{abstract}
We revisit supernova (SN) bounds on a hidden sector consisting of millicharged particles $\chi$ and a massless dark photon. Unless the self-coupling is fine-tuned to be small, rather than exiting the SN core as a gas, the particles form a relativistic fluid and subsequent dark QED fireball, streaming out against the drag due to the interaction with matter. Novel bounds due to excessive energy deposition in the mantle of low-energy SNe can be obtained. The cooling bounds from SN~1987A are unexpectedly not affected in the free-streaming regime. The inclusion of $\chi\bar{\chi}\rightarrow \rm SM$ substantially modifies the constraints in the trapping regime, which can only be treated hydrodynamically. 
Our results can be adapted to generic sub-GeV self-interacting dark sectors.
\end{abstract}

\date{\today}

\maketitle

\begin{figure*}[t!]
    \includegraphics[width=\textwidth]{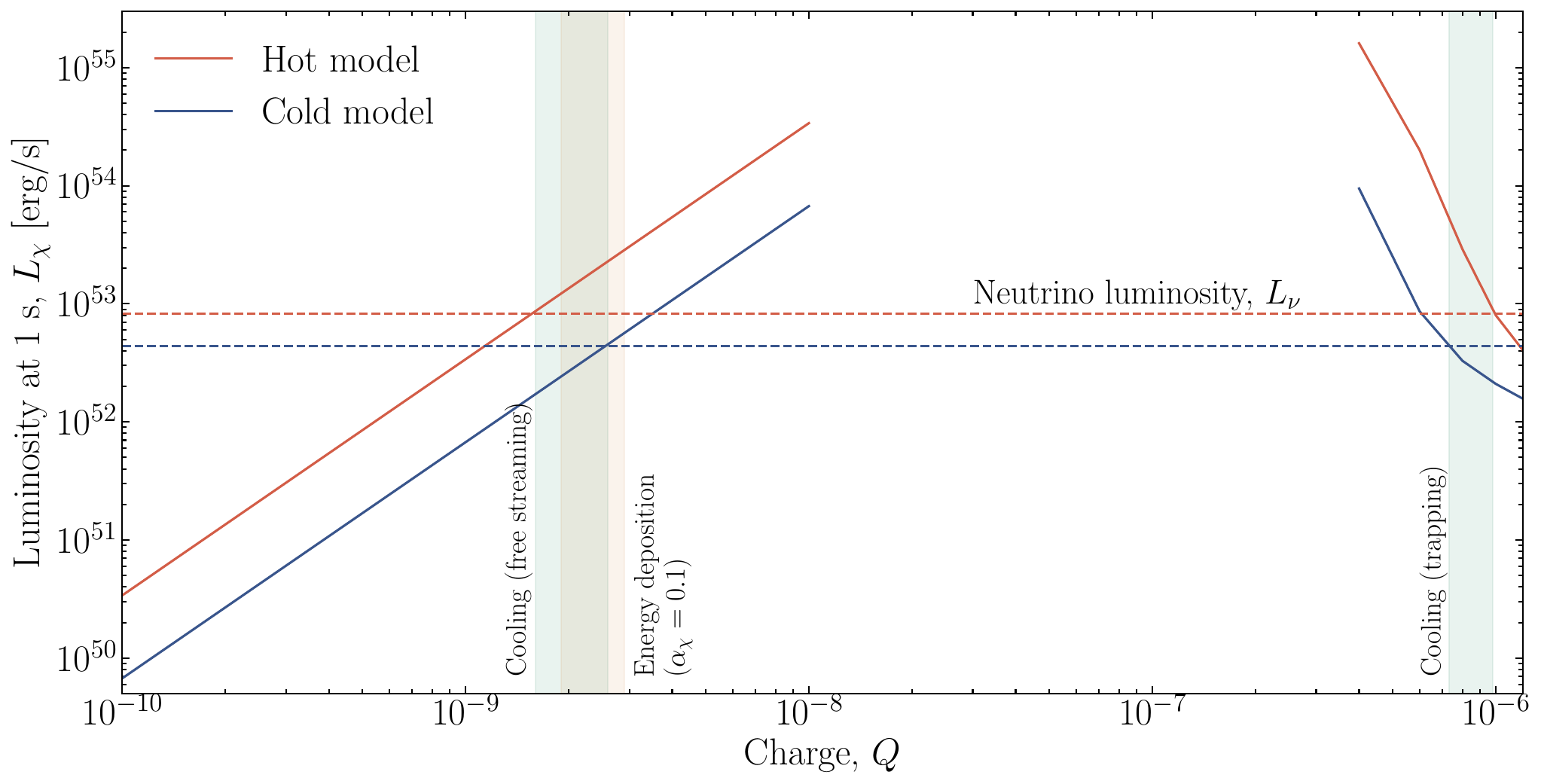}
    \caption{Luminosity of MCPs emitted from the SN core as a function of the MCP charge $Q$. We show as colored bands the bounds from cooling (green) in the free streaming and the regime of trapping from self-interactions $\chi\overline{\chi}\to\gamma$, and the bounds from energy deposition (orange) from self-interactions of MCPs $\chi\overline{\chi}\to\gamma A'$.}\label{fig:fig1}
\end{figure*}

{\bf\textit{Introduction.}}---Supernovae (SNe) and other astrophysical transients have undergone a resurgence  as probes of sub-GeV dark sectors during the last years. Besides bounds on bosons, e.g. axions~\cite{Carenza:2019pxu,Chang:2018rso,Carenza:2020cis,Cavan-Piton:2024ayu}, axion-like particles~\cite{Bollig:2020xdr,Croon:2020lrf,Caputo:2021rux, Ferreira:2022xlw}, and majorons~\cite{Farzan:2002wx, Fiorillo:2022cdq,Akita:2023iwq}, one can draw constraints on self-interacting dark fermions, a paradigmatic example being millicharged particles (MCPs). These are new fermions $\chi$ endowed with a hidden $\rm U(1)_{H}$ symmetry that couples them to a dark photon (DP) mixing kinetically with the SM photon~\cite{Galison:1983pa,Holdom:1986eq} (see also Refs.~\cite{Holdom:1985ag,Dienes:1996zr,Goodsell:2009xc}),
\begin{equation}
    \mathcal{L}\supset-\frac{1}{4}F'_{\mu\nu}F'^{\mu\nu}-\frac{\epsilon}{2}F'^{\mu\nu}F_{\mu\nu}
    +\bar{\chi}(i\gamma^\mu\partial_\mu+g_\chi \gamma^\mu A'_\mu-m_\chi)\chi.
\end{equation}
Here, $g_\chi=\sqrt{4\pi\alpha_\chi}$ is the dark gauge coupling, $m_\chi$ is the MCP mass, the DP is massless, and $\gamma_\mu$ are the Dirac matrices; we use natural units throughout this work. Upon a field redefinition $A_\mu'\rightarrow A_\mu'-\epsilon A_\mu$, the kinetic term is brought into the canonical form, and the MCP is found to have an electric charge $g_\chi \epsilon=e Q \ll e$. Constraints on MCPs have been obtained from laboratory, cosmological, and astrophysical observations~\cite{Jones:1976xy,
Dobroliubov:1989mr,Mohapatra:1990vq,Davidson:1991si,Mitsui:1993ha,
Davidson:1993sj,Prinz:1998ua, Davidson:2000hf,Dubovsky:2003yn,Gies:2006ca,Gninenko:2006fi,
Badertscher:2006fm,
Jaeckel:2009dh,
Jaeckel:2010ni,Diamond:2013oda,
Vogel:2013raa,Moore:2014yba, Vinyoles:2015khy,Chang:2018rso,Magill:2018tbb,Gan:2023jbs,Fung:2023euv}, and their existence has been discussed in the context of observed anomalies of the baryon temperatures at the cosmic dawn~\cite{Barkana:2018lgd,Barkana:2018qrx,Kovetz:2018zan,Munoz:2018pzp,Liu:2019knx,Creque-Sarbinowski:2019mcm}.

Several observables in SNe would be affected by the existence of putative particles $X$ beyond the standard model (SM). The time duration of the neutrino signal detected at Kamiokande~II~\cite{Kamiokande-II:1987idp, Hirata:1988ad, Hirata:1991td, Koshiba:1992yb, Oyama:2021oqp}, the Irvine–Michigan–Brookhaven~\cite{Bionta:1987qt, IMB:1988suc, 1987svoboda}, and the Baksan Underground Scintillation Telescope (BUST) \cite{Alekseev:1987ej, Alekseev:1988gp} experiments from SN~1987A would have been shortened~\cite{Mayle:1987as,Mayle:1989yx,Turner:1987by}. Therefore, the coupling of $X$ to SM particles has to be either small enough to suppress its production (free-streaming regime), or large enough to prevent it from leaving the dense proto-neutron star (PNS) in the core (trapping regime). The traditional criterion is that the additional cooling channel due to the dark sector should not exceed the neutrino luminosity at $1\,\rm s$ after core bounce,
$L_X\lesssim L_\nu(1\rm s)\simeq 3\times 10^{52}\, \rm erg\, s^{-1}$~\cite{Raffelt:2006cw,Caputo:2021rux,Caputo:2022rca}. Additional bounds are obtained, given that core-collapse SNe with low explosion energies (LESNe) have been observed,  by requiring the new particles not to deposit more than $0.1~\rm B$ in the mantle, $1\,\rm B (bethe)=10^{51}\,\rm erg$~\cite{Caputo:2022mah}. (See also Refs.~\cite{Falk:1978kf,Sung:2019xie} for earlier results.)

Our crucial observation is that the existence of self-interactions $\chi\chi\rightarrow\chi\chi$, and similar ones involving the DP, induces self-thermalization and makes the outflow a \textit{relativistic fluid}---rather than a gas---which cannot be treated with kinetic theory, invalidating existing constraints. Energy-momentum conservation of the fluid implies that the free-streaming bounds are quantitatively unaffected. Processes like $\chi\chi\rightarrow\chi\chi$ cannot trap flux, differently from what has been stated recently~\cite{Sung:2021swd}. 
Additionally, we obtain novel constraints from energy deposition in the mantle of LESNe due to processes such as $\chi \chi\rightarrow \chi \chi \gamma$; this is an entirely novel form of energy deposition, driven by self-interactions among, rather than decay of, novel particles---as in, e.g., Refs.~\cite{Falk:1978kf,Caputo:2022mah}. 
Trapping, which can be treated only hydrodynamically (akin to self-interacting neutrinos~\cite{Fiorillo:2023ytr,Fiorillo:2023cas}), is determined by processes $\chi \bar{\chi}\rightarrow \rm SM$, rather than $\chi{\rm SM}\rightarrow\chi\rm SM$ as assumed in previous works~\cite{Davidson:2000hf,Chang:2018rso}. We find that the regions of the parameter space in the large $Q$ regime excluded by the cooling bounds is restricted by roughly one order of magnitude. 

{\bf\textit{Millicharged particle emission.}}---MCPs are dominantly emitted from PNS via plasmon decay. The total energy emitted is obtained e.g. in Ref.~\cite{Raffelt:1996wa}; we summarize these standard formulae in our Supplemental Material (SupM)~\cite{supplementalmaterial}. In order to model the PNS environment, we use as benchmark scenarios
the Garching 1D models SFHo-18.8 and LS220-s20.0 that were evolved with the {\sc Prometheus Vertex} code with six-species neutrino transport~\cite{JankaWeb}, already used for the purpose of particle constraints~\cite{Bollig:2020xdr, Caputo:2021rux,Fiorillo:2022cdq}. With different final NS masses and different equations of state, these models span the extremes of a cold and a hot case, reaching internal $T$ of around 40 vs.\ 60~MeV. For both models, a handle on the MCP properties can be obtained by computing the free-streaming luminosity, namely the total energy emitted in MCPs escaping freely from the PNS. The spectrum of the particles from plasmon decay is reported in standard references~\cite{Raffelt:1996wa}, and we integrate it over the two SN profiles to obtain the MCP luminosity $L_\chi$ and the rate of particle emission $\dot{N}_\chi$; in our SupM we provide fit expressions for their temporal dependence. At $1$~s, where most of the emission happens, we find for the cold (hot) model $L_\chi=0.67 (3.4) \times 10^{52}$~erg/s~$Q_9^2$ and $\dot{N}_\chi=1.8 (5.7) \times 10^{56}$~s$^{-1}$~$Q_9^2$, where $Q_9=Q/10^{-9}$.

The produced MCPs do not freely stream even if $Q$ is very small, because their coupling to the DP allows them to interact with each other. For a luminosity $L_\chi\simeq 10^{52}$~erg/s, assuming a cross section of the order $\sigma_\chi\simeq \alpha_\chi^2/\varepsilon_\chi^2$ as a typical scaling for most processes such as Coulomb scattering or pair annihilation, with $\varepsilon_\chi\simeq 100$~MeV being their average energy, the typical number of collisions at a radius $r$ is
\begin{equation}
    N_\mathrm{coll}\simeq \frac{L_\chi}{4\pi r^2 \varepsilon_\chi}\sigma_\chi r\simeq 3\times 10^{12}\;\alpha_\chi^2 \;\left(\frac{r}{20\;\mathrm{km}}\right)^{-1}.
\end{equation}
For light MCPs and DPs chemical equilibrium can be brought about by dark bremsstrahlung reactions $\chi\chi\to\chi\chi A'$; together with pair annihilation, they break conservation of particle number for all species. The cross section asymptotically scales as $\alpha_\chi^3/m_\chi^2$, but for light MCPs the frequent pair scatterings may interrupt bremsstrahlung radiation if they happen faster than the formation time of a photon with frequency $\omega$, which is of the order of $\varepsilon_\chi^2/m_\chi^2 \omega$, the so-called Landau-Pomeranchuk-Migdal (LPM) effect~\cite{Landau:1953um,Migdal:1956tc}. The bulk of energy emission is for photons with frequency $\omega\simeq \varepsilon_\chi$, so the condition for an efficient LPM effect is $\alpha_\chi^2 n_\chi/m_\chi^2\gg m_\chi^2/\varepsilon_\chi$, where $n_\chi=L_\chi/4\pi r^2 \varepsilon_\chi$ is the MCPs density (notice that, if $m_\chi$ in vacuum is sufficiently light, one should interpret this as the thermal mass of the particle in the plasma, of the order of $m_\chi\sim \sqrt{\alpha_\chi} \varepsilon_\chi$). We can write the number of collisions efficient for bremsstrahlung as
\begin{equation}
    N_\mathrm{coll,br}\simeq \alpha_\chi \mathrm{min}\left[\frac{\alpha_\chi^2 n_\chi}{m_\chi^2},\alpha_\chi \sqrt{\frac{n_\chi}{\varepsilon_\chi}}\right]r,
\end{equation}
so in the LPM regime $N_\mathrm{coll,br}\simeq 6\times 10^{15}\;\alpha_\chi^2$.
Thus, unless a hugely suppressed $\alpha_\chi$ is assumed, these particles will equilibrate rapidly and form a coupled fluid of $\chi$, $\overline{\chi}$, and $ A'$. The subsequent evolution of the fluid is quite different according to whether it freely streams out of the PNS, or it remains trapped. As we will see, self-interactions of MCPs determine the dominant physics in both cases.

{\bf\textit{Free-streaming regime.}}---In the free-streaming regime, the MCP fluid does not interact significantly with the dense PNS, so the cooling rate of the PNS is unaffected by the self-interactions.
While these interactions are very rapid, they cannot trap the particles as claimed in Ref.~\cite{Sung:2021swd}, as they conserve the total momentum of the dark fluid, in the same way as neutrino self-interactions cannot diffusively trap neutrinos~\cite{Dicus:1988jh}. 
In both cases this conclusion is incorrect, since self-interactions render the produced particles a fluid without preventing the outflow from leaving the PNS. 
The problem is thus analogous to the neutrino fluid trapped in the SN core as we discussed in Refs.~\cite{Fiorillo:2023ytr,Fiorillo:2023cas}.

As a first step, we determine the coupling $Q_\mathrm{cool}$ saturating the so-called cooling bound~\cite{Raffelt:2006cw,Caputo:2021rux}. 
Larger MCP emission rates are generally excluded, since they would lead to a rapid cooling of the central PNS, which would shorten the duration of the neutrino burst. This traditional picture has been varyingly questioned in recent years, and the duration of the burst itself has been realized to be sizably altered by convection~\cite{Fiorillo:2023frv}, not included in the original simulations which were used to obtain the cooling criterion. Nevertheless, the cooling bounds still provide a useful first step to investigate the region of parameter space where a non-perturbative impact on SN from the MCP emission can be expected. 
Using the time-dependent luminosity fits obtained earlier, we equal the MCP luminosity at one second with the neutrino luminosity, which is $4.4\;(8.3)\times 10^{52}$~erg/s for the cold (hot) model respectively, finding $Q_\mathrm{cool}=2.6 (1.6)\times 10^{-9}$, in very good agreement with the ones obtained, e.g., in Refs.~\cite{Davidson:2000hf,Chang:2018rso}. 
While the cooling bounds are insensitive to the fluid nature of the MCPs, their fate after emission crucially depends on their hydrodynamical behavior.

After emission, the MCPs quickly equilibrate both thermally and chemically, especially via the bremsstrahlung reaction $\chi\chi\to\chi\chi A'$, with a behavior analogous to the $e^+e^-\gamma$ fireball formed from axion decays in SNe and NS mergers~\cite{Diamond:2023scc,Diamond:2023cto}. The fluid state is determined by its comoving energy density $\rho$ and its radial bulk velocity $v$, with Lorentz factor $\gamma=1/\sqrt{1-v^2}$; its energy flux is $\Phi=4\gamma^2 \rho v/3$. From the conservation of the energy emitted, it follows that at a radius $r$ we have $\Phi=L_\chi/4\pi r^2$.

Outside of the PNS, the fluid freely expands. Its velocity profile is reported, e.g., in Ref.~\cite{Fiorillo:2023cas,Fiorillo:2023ytr}; the fluid escapes with the speed of sound from the emitting surface, while quickly becoming nearly luminal as it moves out. For the problem at hand, there is no sharp decoupling surface, but rather an emission smoothly dropping outside of the PNS. 
One could proceed to find the exact solution of the fluid equations for a given SN profile, but since the uncertainty introduced by using different profiles is much larger than the inaccuracy of the procedure this does not improve much on the precision of the calculation.
Thus, up to factors of order unity, we can conveniently introduce an effective boundary condition at a fixed emission radius $r_e\simeq 14.5$~km; we choose this value to match with the exact numerical solution for the emission obtained at $t=1$~s, when the emission is maximum. With this choice, the asymptotic velocity and density profile reads
\begin{equation}\label{eq:asym_scaling}
    \gamma\simeq \frac{3^{3/4} r}{\sqrt{2}r_e}\simeq 1.6\frac{r}{r_e},\;\rho\simeq \frac{L_\chi r_e^2}{2\pi 3^{3/2}r^4}.
\end{equation}
The value of $r_e$ is chosen such that, for the special snapshot of the cold SN model at $1$~s after bounce, the analytical profile reproduces closely the numerical solution of the fluid equations.
The rest-frame energy density directly determines the temperature, if we use the equation of state for a plasma consisting of two fermion species ($\chi$ and $\overline{\chi}$) with two spin states and vanishing chemical potential, as enforced by the number-changing bremsstrahlung reactions, and one boson species ($ A'$) with two polarization states. For the fermions, up to $10\%$ corrections we can use a Maxwell-Boltzmann distribution; while this approximation is not directly needed at this stage, it will greatly simplify calculations at later stages without affecting sizably any result. With this choice, the comoving energy density reads
\begin{equation}\label{eq:eos}
    \rho=T^4\left(\frac{12}{\pi^2}+\frac{\pi^2}{15}\right),
\end{equation}
where $T$ is the comoving temperature of the fluid. From Eqs.~\eqref{eq:asym_scaling} and~\eqref{eq:eos}, we find
\begin{equation}
    T\simeq 1.6\;\mathrm{MeV}\;\left(\frac{L_\chi}{10^{52}\;\mathrm{erg/s}}\right)^{1/4}\;\frac{r_e}{r}.
\end{equation}

Notice that even at the emitting radius $r_e$ the temperature is entirely independent of the original spectrum with which the MCPs are emitted; even though they are originally produced with $100$~MeV energies, they chemically equilibrate via bremsstrahlung reaching energies of the order of $1$~MeV, similar to Refs.~\cite{Chang:2022gcs,Diamond:2023scc,Diamond:2023cto,Fedderke:2024zzk}. At larger radii, their comoving temperature drops, although their typical lab-frame energy $\gamma T$ remains constant. As the MCPs escape further, they are able to deposit part of their energy onto the material of the progenitor star; this deposited energy cannot be too large, since this would have the dramatic implication of forbidding the existence of LESNe. We thus obtain novel bounds competitive with the cooling based on this complementary observable, entirely dominated by MCP self-interactions.

The dominant channels for energy deposition are annihilations $\chi \overline{\chi}\to \gamma A'$, dark photoproduction $ A' \chi\to\gamma\chi$, and bremsstrahlung $\chi\chi\to\chi\chi\gamma$. In principle, the reaction $\chi\overline{\chi}\to e^+e^-$ could be competitive, but the temperature of the dark plasma is very close to the electron mass and drops with the radius, so it would rapidly become inefficient. We conservatively neglect its effect for this light DP scenario. The rate of energy deposition $dE_\mathrm{dep}/dVdt$ for each of these processes is derived by a scaling of the analogous quantum electrodynamical processes; a derivation of these standard rates is summarized in our SupM, leading to
\begin{equation}
\frac{dE_\mathrm{dep}}{dVdt}=\frac{6\alpha\alpha_\chi Q^2\gamma T^5}{\pi^3}\left(1+\frac{128}{\pi^2}\right)\log\left(\frac{2}{\pi\alpha_\chi}\right).
\end{equation}
At every instant of time, we can integrate over the volume $4\pi r^2 dr$ for $r>r_s$, where $r_s=20$~km is the PNS radius; for $\alpha_\chi=0.1$ we find
\begin{eqnarray}\label{eq:rate_energy_deposition}
    &&\frac{dE_\mathrm{dep}}{dt}\simeq 5.2\times 10^{47}~\mathrm{erg/s}\;\left(\frac{L_\chi}{10^{52}\;\mathrm{erg/s}}\right)^{5/4} Q_9^2.
\end{eqnarray}
Notice that since $L_\chi \propto Q^2$, the deposited energy has a very strong dependence on the coupling constants as $ Q^{9/2}$, meaning that the bounds on $Q$ depend only mildly on uncertainties of order unity in the treatment. We obtain the energy deposition bounds for both the hot and cold model; to do this, we integrate Eq.~\eqref{eq:rate_energy_deposition} over the burst duration, and require it to be smaller than $10^{50}$~erg, as in Ref.~\cite{Caputo:2022mah}. For the cold (hot) model we exclude $Q\gtrsim Q_\mathrm{dep}$, with
\begin{equation}
    Q_\mathrm{dep}\simeq 2.9\;(1.9)\;\times 10^{-9}.
\end{equation}
These bounds are comparable with the cooling bounds obtained earlier, but based on an entirely different observable, adding therefore to the credibility of the exclusion; furthermore, they are sensitive not only to $Q$, but also to the dark coupling $\alpha_\chi$.

\begin{figure}
    \centering
    \includegraphics[width=0.48\textwidth]{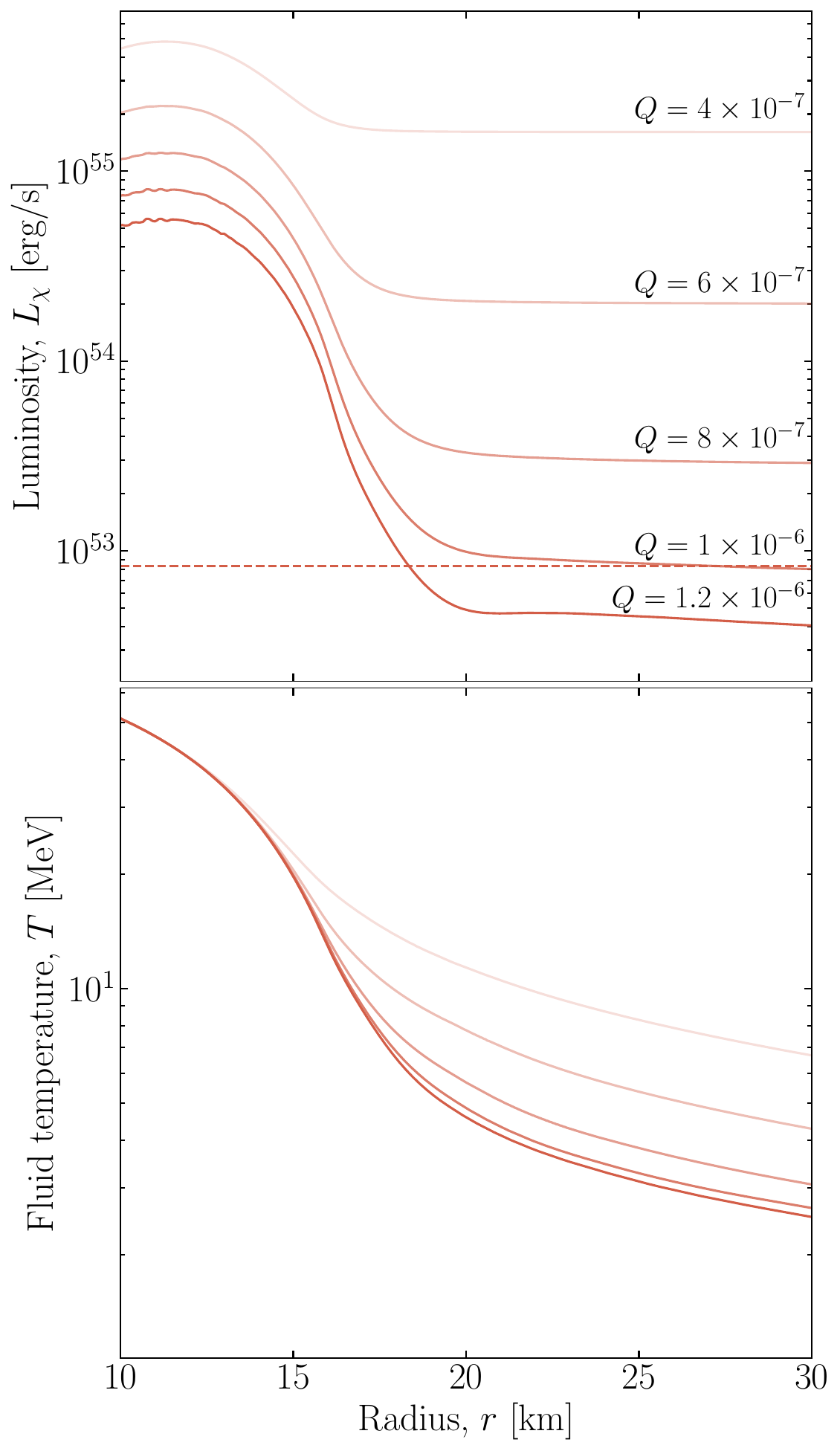}
    \caption{Decoupling of MCPs in the hydrodynamical regime for increasing $Q$. We show the luminosity emitted from a sphere of radius $r$ (top) and the temperature profile (bottom) in the steady state reached at $1$~s for the hot simulation. The neutrino luminosity is shown as a dashed line.}
    \label{fig:fig2}
\end{figure}

{\bf\textit{Trapping regime.}}---As $Q$ grows,  the dark fluid remains trapped in the inner region of the PNS, mostly by processes like $\chi \overline{\chi}\rightarrow \rm SM$.
The MCPs evolution is  ruled by the equations of a relativistic fluid in spherical symmetry, expressing the conservation of energy and momentum
\begin{align}
    \label{eq:fluid_eqs} \nonumber
    &\frac{\partial}{\partial t}\left(\frac{\rho(4\gamma^2+1)}{3}\right)+\frac{1}{r^2}\frac{\partial}{\partial r}\left(\frac{4}{3}\rho\gamma^2 v r^2\right)=\left(\frac{dL_\chi}{dt}\right)_{\mathrm{coal}}, \\\nonumber
    &\frac{\partial}{\partial t}\left(\frac{4}{3}\rho\gamma^2 v\right)+\frac{1}{r^2}\frac{\partial}{\partial r}\left(\frac{\rho}{3}(4\gamma^2v^2+1)r^2\right)-\frac{2\rho}{3 r} \\  &=\left(\frac{d\mathcal{P}}{dVdt}\right)_\mathrm{coal}+\left(\frac{d\mathcal{P}}{dVdt}\right)_\mathrm{Coul},
\end{align}
where $r$ and $t$ are the radius and time, and the right-hand terms describe the energy and momentum exchanged per unit volume and time with the PNS matter via the reaction of coalescence $\chi\overline{\chi}\to\gamma$ and Coulomb scattering $\chi N\to\chi N$. The energy exchanged via Coulomb scattering is negligible since the nuclei are much heavier than the plasma temperature. The concrete form of these terms must be obtained by averaging the microscopic interaction rate over the MCPs distribution, which is maintained thermalized at all times by the dark self-interactions. We perform this calculation in detail in our SupM. We include plasmon coalescence as the only $\chi\chi\to $SM process. For $\alpha_\chi\gtrsim\alpha$ photoproduction, pair annihilation, and bremsstrahlung may be more efficient, so the onset of MCP trapping could start at yet smaller $Q$.

At every simulation snapshot, the MCP fluid moves with the speed of light and is able to relax to a stationary state, where in the inner regions is in equilibrium with the PNS medium, while at outer regions it decouples and escapes freely. For growing $Q$, it remains coupled up to outer layers where the temperature is lower, at some point leading to an emitted thermal luminosity smaller than the neutrino one, not anymore excluded from the cooling argument. While this is a well-known feature, in the special case of MCPs the conventional approach of using a blackbody emission from a characteristic particle sphere is not valid, since the MCPs are strongly coupled to each other. We must resort to the fluid Eqs.~\eqref{eq:fluid_eqs} to describe the steady MCP emission. In practice, we consider the SN profile at $1$~s, where the MCP emission peaks, and we evolve the fluid equations starting from no MCPs until they reach the stationary state.

Fig.~\ref{fig:fig2} shows the steady emission of MCPs for increasing values of $Q$ for our hot model. The picture looks remarkably similar to the temporal decoupling in the early Universe, except that it is a spatial decoupling; at sufficiently small radii the MCP fluid is always in equilibrium with the medium, while it decouples when the energy exchanged in a light-crossing time becomes negligible compared with the energy of the MCP fluid. At this point, the fluid relaxes into the freely expanding solution discussed earlier, with the temperature $T\propto r^{-1}$. The position of the decoupling radius is essentially set by the strength of the self-interaction $\chi\overline{\chi}\to\gamma$, which is the dominant reaction at a radius $r\sim 20$~km. Coulomb interactions are marginally competitive in establishing momentum exchange, but cannot maintain thermal and chemical equilibrium; we have checked that removing the Coulomb term from the fluid equations only marginally changes the obtained bounds. The key quantity describing the cooling rate of the PNS is the energy passing through a sphere of radius $r$ per unit time, shown in the top panel of Fig.~\ref{fig:fig2}. At inner radii, the fluid is coupled to the PNS material, so the energy flux is driven by the gradients in the temperature profile of the SN, while after decoupling the fluid escapes freely and thus the energy passing through outer spheres is conserved; its value is exactly the non-standard cooling rate of the PNS, which by virtue of the cooling criterion should be smaller than the neutrino luminosity at the same time. This allows us to identify the trapping regime above which no robust exclusion holds.

{\bf\textit{Discussion and outlook.}}---We revisited the emission of self-interacting particles from SNe focusing on the case of MCPs. We have shown that the evolution of the outflow is entirely driven by self-interactions both in the free-streaming and in the trapping regime. In the small $Q$ regime we have found novel LESNe calorimetric constraints. For heavier DPs, dark bremsstrahlung reactions might become kinematically inhibited before reaching chemical  equilibrium, an orthogonal regime worth exploring in future work. 
In the trapping regime, the bounds stop at $Q\sim 10^{-6}$, nearly an order of magnitude smaller than the latest estimate~\cite{Chang:2018rso}. 
MCPs do not decouple individually, but rather as a fluid, making the conventional kinetic treatment~\cite{Chang:2018rso,Caputo:2022rca} inapplicable; the same is true for any dark sector with large enough self-couplings.

%MCPs provide an intriguing example of new physics where self-interactions decide on their fate and their constraints in SNe. 
If MCPs were not a fluid, their gyroradius at a typical energy $\varepsilon_\chi\simeq 100$~MeV would be of the order of $r_g\simeq 3\times 10^{14}$~cm~$B_{\rm G}^{-1} Q_9^{-1}$, where $B_{\rm G}$ is the magnetic field in Gauss. For $B\sim 10^{10}$~G, $r_g\ll r_s$ and the particles might potentially get trapped. Without self-interactions, this effect has never been included, and might hinder the cooling bounds. For self-interacting MCPs, which we have shown to be fluid, oppositely charged particles isotropize much before they can get deflected. Cooling bounds are robust against magnetic field uncertainties \textit{only} because MCPs are fluid.
%Since the MCPs behave as a fluid their mean free path for self-interactions is much shorter than the gyroradius, so MCPs with opposite charges isotropize much before they can get deflected. Therefore, the constraint from SN~1987A cooling exists \textit{only} because MCPs are a fluid.
The magnetic field might still induce electric currents in the neutral dark fluid, deflecting the fluid by Lorentz forces, but the corresponding conductivity is suppressed both by the short mean free path and by $Q^2$, leading to negligible effects. Thus, the fluid outflow has a direct impact even on the standard cooling bounds, making them robust to systematic uncertainties on magnetic fields. We leave the case of MCPs not featuring a DP, i.e. leaving as a gas influenced by magnetic fields, for a forthcoming publication.

Following our approach, SN bounds on novel particles such as ELDER~\cite{Kuflik:2015isi}, SIMP~\cite{Hochberg:2014dra}, and other dark sectors (see e.g.~\cite{Boddy:2014yra}) can be revisited by appropriately taking into account self-interactions. This is important in light of the ever-growing interest in sub-GeV dark matter searches~(see e.g.~\cite{Essig:2013lka,Essig:2022dfa,Zurek:2024qfm} and references therein).

\textbf{\textit{Acknowledgments}}---We thank Georg Raffelt for comments on a draft of this paper.
DFGF is supported by the Villum Fonden under Project No. 29388 and the European Union’s Horizon 2020 Research and Innovation Program under the Marie Sklodowska-Curie Grant Agreement No. 847523
``INTERACTIONS.'' This work is supported by the Italian MUR Departments of Excellence grant 2023- 2027 ``Quantum Frontiers'' and by Istituto Nazionale di Fisica Nucleare (INFN) through the Theoretical Astroparticle Physics (TAsP) project.

\onecolumngrid
\appendix

\clearpage

\setcounter{equation}{0}
\setcounter{figure}{0}
\setcounter{table}{0}
\setcounter{page}{1}
\makeatletter
\renewcommand{\theequation}{S\arabic{equation}}
\renewcommand{\thefigure}{S\arabic{figure}}
\renewcommand{\thepage}{S\arabic{page}}

\begin{center}
\textbf{\large Supplemental Material for the Letter\\[0.5ex]
{\em Self-Interacting Dark Sectors in Supernovae Can Behave as a Relativistic Fluid}}
\end{center}

In this Supplemental Material, we provide an explicit derivation for the rates of the processes discussed in the main text. Specifically, we compute the rate of coalescence of millicharged particles (MCPs) to plasmons, and obtain the rate of energy and momentum transfer between the standard model (SM) fluid and the MCP fluid via this reaction. We discuss the rate of energy deposition into the SM plasma of the dark fluid via dark electrodynamical reactions of the form $\chi\overline{\chi}\to\gamma A'$, $ A'\chi\to\gamma\chi$, $\chi\chi\to\chi\chi\gamma$. We obtain a collisional term for Coulomb reactions $\chi N \to \chi N$ in the SM plasma, and use it to derive the rate of momentum transfer between the SM fluid and the MCP fluid via Coulomb scatterings. Finally, we compare our bounds with the bounds on non-self-interacting MCPs previously used in the literature.

\bigskip

\section{A.~Coalescence of millicharged particles}

In Supernova (SN) cores, the frequency $\omega$ and wave number $k$ of the electromagnetic field excitations do not satisfy the vacuum dispersion relation $\omega=k$, due to refraction on the dense matter. The deviations from the light-like dispersion relation is roughly measured by the plasma frequency $\omega_{\rm P}$, and the corresponding excitations are dubbed plasmons, which can have longitudinal or transverse polarization. For simplicity, we will assume throughout the hierarchy $m_\chi\ll \omega_{\rm P}\ll T$, which guarantees that both MCPs and plasmons are usually relativistic. In these conditions, the dispersion relation of plasmons can be found from the implicit form $\omega^2=k^2+\pi_L(\omega,k)$, where $\omega$ and $k$ are the plasmon energy and momentum; using the ultra-relativistic form for the polarization function $\pi_L$ (see Eq.~4.38 in Ref.~\cite{Raffelt:1996wa})
\begin{equation}
    \omega^2=k^2+\frac{3\omega_{\rm P}^2}{2}\left(1-\frac{k^2}{\omega^2}\right)\left[-2+\log\left(\frac{4}{1-\frac{k^2}{\omega^2}}\right)\right],
\end{equation}
which admits the implicit solution
\begin{equation}
    \frac{k^2}{\omega^2}=1-4 e^{-2-\frac{2\omega^2}{3\omega_{\rm P}^2}}.
\end{equation}
Therefore relativistic longitudinal plasmons are exponentially close to the light cone, and in turn their decay is kinematically suppressed exponentially. In this work we will consider only transverse plasmons, whose ultra-relativistic dispersion relation is $\omega^2\simeq k^2+3\omega_{\rm P}^2/2$. Their wavefunction is weakly renormalized in the ultra-relativistic regime; see Ref.~\cite{Raffelt:1996wa} for a textbook treatment.

The dominant processes by which MCPs can be produced is plasmon decay, $\gamma\to \overline{\chi}\chi$. Previous authors have also included additional processes, such as Compton $\gamma e^-\to \overline{\chi}\chi e^-$ and nucleon-nucleon bremsstrahlung. However, in all these processes the decay of an intermediate plasmon always happens; if this plasmon is on-shell, considering these processes separately leads to a double counting. In reality, these additional processes mainly endow the plasmon propagator with a finite absorption width. Provided this width is much smaller than the plasmon frequency, which is generally the case since they are suppressed by at least one power of the fine structure constant $\alpha$, they can be neglected---i.e., plasmons can be considered particles with a narrow width. For this reason, we consider here only the on-shell decay of plasmons; were these processes to be included, they would enter as a finite width for the delta function enforcing the plasmons to be on-shell.

The rate of absorption of a MCP $\chi$ with momentum $\bp$ and energy $p=|\bp|$, due to its coalescence with MCP $\chi\overline{\chi}\to\gamma$, can be written as
\begin{equation}\label{eq:rate}
    \left(\frac{\partial f_\bp}{\partial t}\right)_\mathrm{abs}=-\frac{f_\bp}{2p}\int \frac{d^3\bp'}{(2\pi)^3}\frac{2}{2p'} \frac{d^3\bk}{(2\pi)^3 2\omega_\bk}(2\pi)^4\delta^{(3)}(\bp+\bp'-\bk) \delta(p+p'-\omega_\bk) f_{\bp'}(1+f^\gamma_\bk)|\mathcal{M}|^2.
\end{equation}

Here $f_\bp$ is the phase-space distribution function of the MCP $\chi$ with momentum $\bp$; notice that we assume throughout that the distributions of $\chi$ and $\overline{\chi}$ are exactly equal, an obvious conclusion of the charge-symmetric nature of the problem. Further, $\bp'$ is the momentum of $\overline{\chi}$ with which $\chi$ is coalescing; $\bk$ and $\omega_\bk$ are the momentum and energy of the produced plasmon; $f^\gamma_\bk$ is the plasmon phase-space distribution function; $|\mathcal{M}|^2$ is the spin-averaged matrix element summed over the polarization states of the transverse plasmon. The factor $2$ in the numerator accounts for the two helicity states of the target MCP $\overline{\chi}$. By the law of detailed balance this process entails also the production of MCPs via plasmon decay,
\begin{equation}\label{eq:decay_rate}
    \left(\frac{\partial f_\bp}{\partial t}\right)_\mathrm{prod}=\frac{1}{2p}\int \frac{d^3\bp'}{(2\pi)^3}\frac{2}{2p'} \frac{d^3\bk}{(2\pi)^3 2\omega_\bk}(2\pi)^4\delta^{(3)}(\bp+\bp'-\bk) \delta(p+p'-\omega_\bk) f^\gamma_{\bk}(1-f_\bp)(1-f_{\bp'})|\mathcal{M}|^2.
\end{equation}
Even in chemical equilibrium, since $\chi$ and $\overline{\chi}$ are produced in pairs, their chemical potential would be $0$, and so the degeneracy factors are under no conditions very different from $1$ for $p,p'\sim T$. Thus, they can be neglected.

We are going to assume temporarily that the absorption is dominated by reactions in which relativistic plasmons are produced in the limit $\omega_{\rm P}\ll T$, an assumption that will be verified by the results. Under these assumptions, as discussed in the main text, the dispersion relation for the plasmon is
\begin{equation}
    \omega_\bk^2=k^2+m_\gamma^2,
\end{equation}
with $m_\gamma^2=3\omega_{\rm P}^2/2$. To lowest order in $m_\gamma/T$, we can neglect the plasmon mass everywhere except in the phase-space integration, where obviously an exactly massless plasmon could not be produced by coalescence. In particular, the matrix element in general is given by (see, e.g., Eq.~(6.77) in Ref.~\cite{Raffelt:1996wa})
\begin{equation}
    |\mathcal{M}|^2=4\pi\alpha Q^2 p_\beta p'_\alpha \sum_{I=1}^{2}\left[g^{\alpha\beta}+2 Q_I^\alpha  Q_I^\beta\right],
\end{equation}
where $ Q_I^\alpha$ is the polarization four-vector of the produced plasmon in the $I$-th state of polarization. Using the on-shell conditions enforced by the delta functions, one can reduce this matrix element to the form
\begin{equation}
    |\mathcal{M}|^2=2\pi\alpha Q^2m_\gamma^2\frac{2(p^2+{p'}^2)-m_\gamma^2}{(p+p')^2-m_\gamma^2}.
\end{equation}

In the limit of very small $m_\gamma\ll p, p'$ we can simplify this to 
\begin{equation}
    |\mathcal{M}|^2=4\pi\alpha Q^2m_\gamma^2\frac{p^2+{p'}^2}{(p+p')^2}.
\end{equation}

In Eq.~\eqref{eq:rate} we can enforce momentum conservation, leading to the final expression
\begin{equation}
    \left(\frac{\partial f_\bp}{\partial t}\right)_{\mathrm{abs}}=-\frac{\alpha Q^2 m_\gamma^2 f_\bp}{4\pi}\int d^3 \bp' f_{\bp'}(1+f^\gamma_{\bp+\bp'}) \frac{p^2+{p'}^2}{p p'(p+p')^3} \delta(p+p'-\sqrt{|\bp+\bp'|^2+m_\gamma^2}).
\end{equation}
In the limit $m_\gamma\ll T$ that we are taking, the integral can be simplified further. Energy conservation implies that the angle $\theta_{\bp\bp'}$ between $\bp$ and $\bp'$ should be~\footnote{We neglect here the MCP mass, which is supposed to be such that $m_\chi\ll m_\gamma$. Notice however that for longitudinal plasmons the equivalent photon mass $m_\gamma$ is exponentially suppressed with momentum as $e^{-2k^2/3\omega_{\rm P}^2}$, so this inequality is generally not valid. Indeed, we neglect here longitudinal plasmons entirely, under the assumption that $m_\chi\gtrsim T e^{-T^2/3\omega_{\rm P}^2}$.} 
\begin{equation}
    \cos\theta_{\bp\bp'}=1-\frac{m_\gamma^2}{2p p'}.
\end{equation}
Thus, $\theta_{\bp\bp'}$ is of the order $m_\gamma/T$ for most collision events. If we now separate the dependence of the distribution function on the module $p$ and the direction $\bn$ of the momentum $f_\bp=f(p,\bn)$, provided that $f(p,\bn)$ does not change rapidly over angles of order $m_\gamma/T$, we can evaluate $f_{\bp'}=f(p',\bn')\simeq f(p',\bn)$. Furthermore, since plasmons remain in thermal equilibrium throughout, we can simply replace $f^\gamma_{\bp+\bp'}=f_{\rm BE}(p+p')$, where $f_{\rm BE}$ is the Bose-Einstein distribution function. 
Performing the integral over the angle $\bn'$ analytically we are finally led to 
\begin{equation}
    \left(\frac{\partial f(p,\bn)}{\partial t}\right)_{\mathrm{abs}}=-\frac{\alpha Q^2m_\gamma^2f(p,\bn)}{2p^2}\int dp' f(p',\bn)(1+f_{\rm BE}(p+p'))\frac{p^2+{p'}^2}{(p+p')^2}.
\end{equation}
Notice that these relations are meant to be used only in the range $p\gg m_\gamma, m_\chi$, so the divergence at small $p$ is obviously not physical. Since the MCPs form a fluid, this divergence does not produce any formal problem; the total energy and momentum of the fluid are always obtained by multiplication of the phase-space volume $p^2 dp$, so the low-energy MCPs give negligible contribution to any hydrodynamical quantity.

The full collisional integral originating from the coalescence process can now be written accounting also for the inverse process as
\begin{equation}\label{eq:coll_coal}
\left(\frac{\partial f(p,\bn)}{\partial t}\right)_\mathrm{coal}=\frac{\alpha Q^2 m_\gamma^2}{2p^2}\int dp'\frac{1}{1-e^{-\frac{p+p'}{T_\gamma}}}\frac{p^2+{p'}^2}{(p+p')^2}\left[e^{-\frac{p+p'}{T_\gamma}}-f(t,\br,p,\bn)f(t,\br,p',\bn)\right],
\end{equation}
where $T_\gamma$ is the photon temperature, in general different from the MCPs temperature.

To verify the correctness of this equation, we may use it in the free streaming regime, where the last term is negligible, to obtain the luminosity emitted per unit volume
\begin{equation}
    \frac{dL_\chi}{dV}=\int \frac{d^3\bp}{(2\pi)^3}2\int dp' (p+p') \frac{\alpha  Q^2 m_\gamma^2}{2p^2}\frac{p^2+{p'}^2}{(p+p')^2}\frac{1}{e^{\frac{p+p'}{T_\gamma}}-1},
\end{equation}
where we multiply by a factor $p+p'$ which is the energy carried by the pair. The integral is easily done and gives
\begin{equation}\label{eq:free_streaming_lum}
    \frac{dL_\chi}{dV}=\frac{2\alpha Q^2 m_\gamma^2 T_\gamma^3 \zeta(3)}{3\pi^2},
\end{equation}
while the total number of emitted particles is
\begin{equation}
    \frac{dN_\chi}{dVdt}=\frac{\alpha Q^2 m_\gamma^2 T_\gamma^2}{9}.
\end{equation}
We can compare this with the luminosity obtained using the decay rate of the plasmon reported in Ref.~\cite{Raffelt:1996wa}
\begin{equation}
    \Gamma(\omega)=\frac{\alpha  Q^2 m_\gamma^2}{3\omega},
\end{equation}
which leads for the luminosity to the result
\begin{equation}
    \frac{dL_\chi}{dV}=\int \frac{\omega^2 d\omega}{\pi^2} \omega \Gamma(\omega) f_{\rm BE}(\omega)
\end{equation}
which when integrated gives the same result as Eq.~\eqref{eq:free_streaming_lum}.

\begin{figure}
    \centering
    \includegraphics[width=0.5\textwidth]{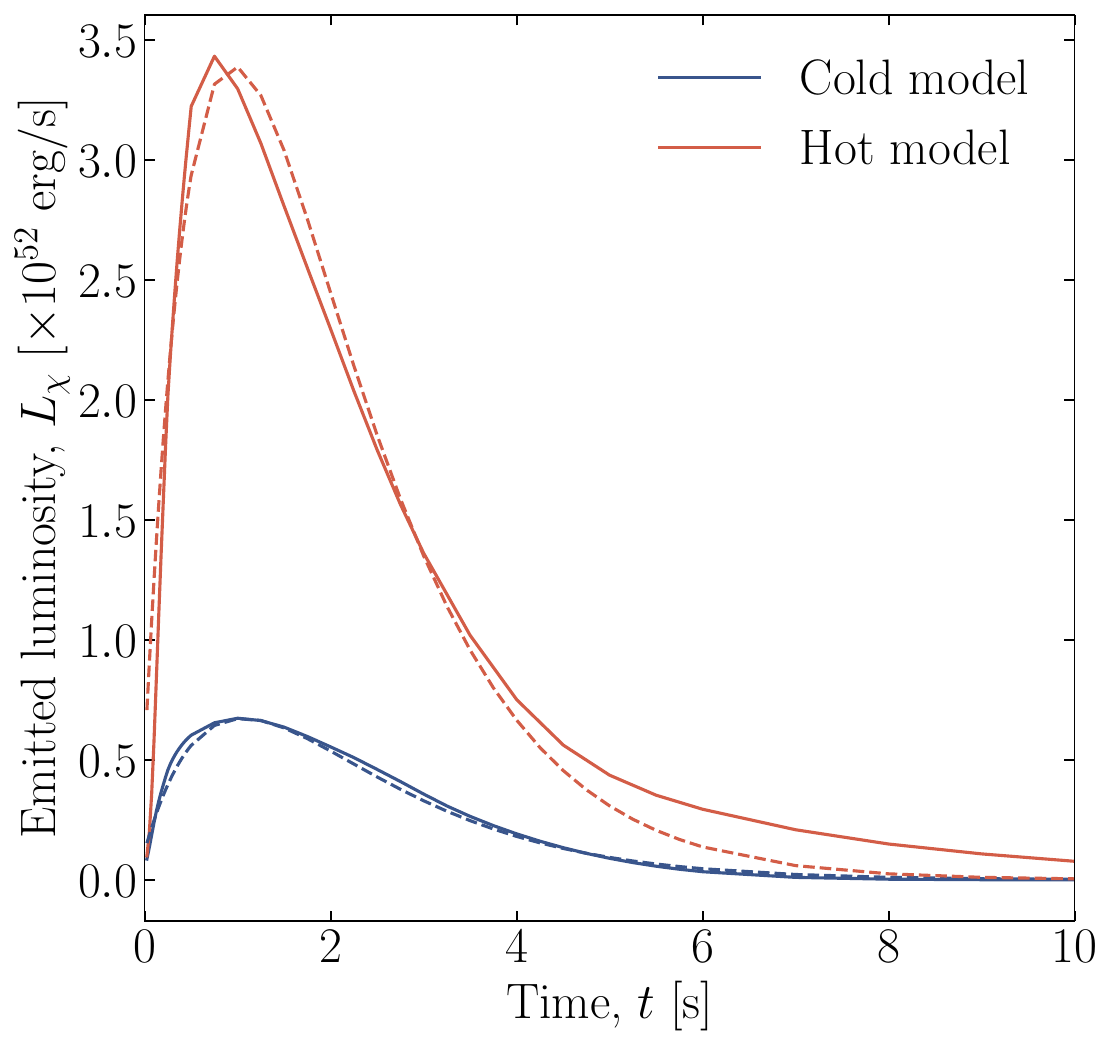}
    \caption{ Comparison of the instantaneous luminosity from the numerical SN models (solid), compared with the fit expressions provided in the text (dashed), as a function of time, for a value of the charge $Q=10^{-9}$. }
    \label{fig:lum_fit}
\end{figure}

We have integrated these expressions over the temperature and density profiles of the hot and cold model introduced in the text. We find that an excellent fit to the time-dependent luminosity for the cold model is
\begin{equation}
    L_\chi\simeq 1.5\times 10^{52}\; \mathrm{erg/s}\;  Q_9^2\;(t+0.08\;\mathrm{s})\;\mathrm{exp}\left[-\frac{t}{1.14\;\mathrm{s}}\right],
\end{equation}
while for the hot model
\begin{equation}
    L_\chi\simeq 8.6\times 10^{52}\; \mathrm{erg/s}\;  Q_9^2\;(t+0.06\;\mathrm{s})\;\mathrm{exp}\left[-\frac{t}{1.01\;\mathrm{s}}\right],
\end{equation}
where $ Q_9= Q/10^{-9}$.  The comparison of these fit expressions with the exact numerical luminosity is shown in Fig.~\ref{fig:lum_fit}.

In the trapping regime, MCPs thermalize via their dark self-interactions much faster than coalescence or any other interaction with the Standard Model. Thus, the collisional term in Eq.~\eqref{eq:coll_coal} acts as an exchange of energy and momentum for the dark fluid with the protoneutron star (PNS) material. In principle also the number of $\chi$ is changed by the coalescence process, but the dark interactions change the number of $\chi$ much more rapidly via dark bremsstrahlung, adjusting it to the conditions of chemical equilibrium.
Let us thus determine the rate of energy exchange with the medium in the limit where the temperature of the dark plasma $T$ is very close to the temperature of the medium $T_\gamma$, which is relevant for the trapping regime. The energy given by the medium to the dark plasma is
\begin{equation}
    \frac{dL_\chi}{dV}=\int \frac{d^3\bp}{(2\pi)^3}2\int dp' (p+p') \frac{\alpha  Q^2 m_\gamma^2}{2p^2}\frac{p^2+{p'}^2}{(p+p')^2}\frac{1}{e^{\frac{p+p'}{T_\gamma}}-1}\left[1-\exp\left[\frac{p+p'}{T_\gamma}-\gamma\frac{p+p'}{T}(1-v x)\right]\right],
\end{equation}
where $x$ is the cosine of the angle between the particle direction $\bn$ and the bulk velocity of the fluid. Expanding the exponential for small $\delta T=T-T_\gamma$ and small velocities, we finally find after performing the integral that
\begin{equation}
    \frac{dL_\chi}{dV}=-\frac{\alpha Q^2 m_\gamma^2\pi^2 T_\gamma^2 \delta T}{45},
\end{equation}
which therefore act as a temperature relaxation term trying to maintain the temperature of the dark fluid equal to the photon temperature. We can in the same way determine the momentum given to the MCP fluid per unit time and volume as
\begin{equation}
   \frac{d\boldsymbol{\mathcal{P}}}{dVdt}=-\frac{\alpha Q^2 m_\gamma^2\pi^2 T_\gamma^3 \bv}{135}.
\end{equation}

\section{B.~Rate of energy deposition from dark fluid}

The dark fluid can deposit energy into the SM plasma by various electrodynamical processes in which one of the dark photons is replaced by a SM photon; these include the pair annihilation $\chi\overline{\chi}\to\gamma A'$, the photoproduction $ A'\chi\to\gamma\chi$, and bremsstrahlung $\chi\chi\to\chi\chi\gamma$. An evaluation of the rates for these processes to great precision would presumably be a complex task, especially since the masses of the MCPs and the DPs would be dominated by thermal effects in the dense dark plasma. A detailed thermal treatment is probably unwarranted, since our main aim is simply to show that the bounds from energy deposition are comparable with the bounds from PNS cooling. Thus, in this section we derive the order of magnitude of the energy deposited by these processes. We assume that the dark plasma has thermalized via processes unsuppressed by powers of $Q$, discussed in the main text, and so is defined by a temperature $T$. In such a state, the MCPs have a thermal mass~\cite{Braaten:1991hg}
\begin{equation}
    m_\chi=\sqrt{\frac{\alpha_\chi\pi}{2}}T;
\end{equation}
in the spirit of obtaining simple approximations with the correct order of magnitude, we will assume $m_\chi\ll T$ even for $\alpha_\chi=0.1$.

MCP annihilation $\chi\overline{\chi}\to\gamma A'$ is analogous to electron-positron annihilation in two photons, with the difference that the coupling to the photon and the DP are respectively $\propto  Q\sqrt{\alpha}$ and $\propto \sqrt{\alpha_\chi}$. Thus, we can immediately write the cross section for the process, in the limit of ultrarelativistic $\chi$~\cite{Akhiezer:1986yqm}
\begin{equation}\label{eq:sigma_ann}
    \sigma_{\mathrm{ann}}=\frac{\alpha\alpha_\chi  Q^2 \pi}{\varepsilon_0^2}\log\left[\frac{2\varepsilon_0}{m_\chi}\right],
\end{equation}
where $\varepsilon_0$ is the energy of $\chi$ in the center-of-mass system of the collision. We will denote by tilde the quantities in the frame comoving with the fluid, and without tilde the quantities in the laboratory frame. For two particles $\chi$ and $\overline{\chi}$ with energy $\tilde{p}_1$ and $\tilde{p}_2$ respectively, and angle between the two momenta $\tilde{x}=\cos\tilde{\theta}$, we have $\varepsilon_0=\sqrt{\tilde{p}_1 \tilde{p}_2(1-\tilde{x})/2}$. In the comoving frame, after averaging over the isotropically distributed incoming directions, the energy transferred to the Standard Model photon is $(\tilde{p}_1+\tilde{p}_2)/2$, as expected from symmetry. In the laboratory frame, this corresponds to a deposited energy $\gamma(\tilde{p}_1+\tilde{p}_2)/2$, after averaging over the isotropic direction of the photon. Therefore, the rate of energy deposition per unit volume and time is
\begin{equation}
    \left(\frac{dE_\mathrm{dep}}{dVdt}\right)_{\mathrm{ann}}=\int f(\tilde{p}_1) \frac{2 d^3\tilde{\bp}_1}{(2\pi)^3} f(\tilde{p}_2) \frac{2 d^3\tilde{\bp}_2}{(2\pi)^3} (1-\tilde{x}) \frac{\gamma(\tilde{p}_1+\tilde{p}_2)}{2} \sigma_\mathrm{ann};
\end{equation}
the factor $1-\tilde{x}$ accounts for the relative velocity between the two particles. The distribution function must be taken as the Maxwell-Boltzmann function with temperature $T$, consistently with our description in the main text.

In the cross section, the logarithm is slowly varying and we can approximately replace it with $\log(2\varepsilon_0/m_\chi)\simeq \log(T/m_\chi)$ up to factors of order unity; since the emissivity is proportional to a large power of $ Q$, this makes little difference on the bounds we obtain in the main text. With this approximation, we only need to do the integrals over the module of $\tilde{\bp}_1$ and $\tilde{\bp}_2$, since the factor $1-\tilde{x}$ cancels with the analogous factor in the denominator of $\sigma_\mathrm{ann}$. The remaining integrals are easily done leading to the final expression
\begin{equation}
    \left(\frac{dE_\mathrm{dep}}{dVdt}\right)_{\mathrm{ann}}=\frac{4\alpha\alpha_\chi  Q^2 \gamma T^5}{\pi^3}\log\left(\frac{T}{m_\chi}\right).
\end{equation}

Regarding photoproduction, the total cross section has actually the same form as Eq.~\eqref{eq:sigma_ann} in the ultrarelativistic limit. Since the number of degrees of freedom for $ A'$, $\chi$, and $\overline{\chi}$ are all equal, if we approximate all of their distributions by Maxwell-Boltmann it
follows that the energy deposited by the two reactions $ A'\chi\to\gamma\chi$ and $ A'\overline{\chi}\to\gamma\overline{\chi}$ are all equal to the annihilation rate, so we just need to multiply by a factor $3$ to obtain the total rate of energy deposition by pair annihilation and photoproduction.

Finally, we consider energy deposition from bremsstrahlung $\chi\chi\to\chi\chi\gamma$; the cross section for energy loss is given, e.g., in Eq.~(9) of Ref.~\cite{ALEXANIAN:1968fdp}, so that the average energy lost in the comoving frame of the fluid by two colliding particles  with energies $\tilde{p}_1$ and $\tilde{p}_2$ is
\begin{equation}
    \int \frac{d\sigma_\mathrm{brem}}{d\omega}\omega d\omega=\Phi_\mathrm{brem}(\tilde{p}_1+\tilde{p}_2)=\frac{8\alpha_\chi^2\alpha Q^2}{m_\chi^2}\log\left(\frac{2\varepsilon_0}{m_\chi}\right) (\tilde{p}_1+\tilde{p}_2),
\end{equation}
where again $\varepsilon_0$ is the energy of $\chi$ in the center-of-mass rest frame. Hence, the total energy deposited is
\begin{equation}
    \left(\frac{dE_\mathrm{dep}}{dVdt}\right)_{\mathrm{brem}}=\int f(\tilde{p}_1) \frac{4 d^3\tilde{\bp}_1}{(2\pi)^3} f(\tilde{p}_2) \frac{4 d^3\tilde{\bp}_2}{(2\pi)^3} (1-\tilde{x}) \gamma(\tilde{p}_1+\tilde{p}_2) \Phi_\mathrm{brem};
\end{equation}
notice that we must now account for $4$ degrees of freedom for the $\chi$ particles, since both $\chi$ and $\overline{\chi}$ can participate to the process. Using again Maxwell-Boltzmann distributions for the integration, with the usual simplification that $\log(2\varepsilon_0/m_\chi)\simeq \log(T/m_\chi)$, we finally find
\begin{equation}
    \left(\frac{dE_\mathrm{dep}}{dVdt}\right)_\mathrm{brem}\simeq \frac{768\alpha_\chi^2\alpha Q^2 \gamma T^7}{\pi^4 m_\chi^2}\log\left(\frac{T}{m_\chi}\right).
\end{equation}
We can now replace $m_\chi=\sqrt{\frac{\pi \alpha_\chi}{2}}T$ to find that the bremsstrahlung emission has also the same form as the pair annihilation emission with a different numerical coefficient in front. In obtaining the bremsstrahlung emission, we have neglected the Landau–Pomeranchuk–Migdal (LPM) effect. The coherence length for a photon with typical energy $T$ is of the order of $\ell_\mathrm{coh}\simeq T/m_\chi^2 \sim 1/\alpha_\chi T$, while the typical length between two collisions efficient for bremsstrahlung is $m_\chi^2/\alpha_\chi^2 n_\chi\sim 1/\alpha_\chi T$, where $n_\chi \sim T^3$ is the number density. Since the two lengths are comparable, we are close to the regime in which the LPM effect can be relevant.
Since a detailed treatment of the LPM effect beyond orders of magnitude would anyway be very complex, we neglect it here, with the understanding that our expressions are correct in order of magnitude.

Putting everything together, we find that the total energy deposited via all the above processes is
\begin{equation}
    \frac{dE_\mathrm{dep}}{dVdt}\simeq \left[1+\frac{128}{\pi^2}\right]\frac{12\alpha\alpha_\chi Q^2\gamma T^5}{\pi^3}\log\left(\sqrt{\frac{2}{\pi\alpha_\chi}}\right).
\end{equation}

\section{C.~Coulomb scattering on protons}

In the trapping region, not only self-interactions, but also Coulomb scatterings on protons happen rapidly, and must be included in the transport treatment. The impact of Coulomb interactions on the trapping of the species is tricky, since they conserve the number of MCPs, and only allow for a slow energy transfer with the medium, due to the nucleons being heavy compared to the MCP temperature. In this section, we derive the collisional integral for Coulomb scattering. In the general form, this is quite a complex task beyond our purposes. However, this problem is usually simplified in plasma physics by the observation that most collisions are strongly forward in the center-of-mass frame, due to the divergence of the Coulomb cross section for small-angle scatterings. In this regime, the collisional integral can be given a Fokker-Planck form~\cite{planck1917satz,fokker1914mittlere} that we now derive, in the same way as the Landau collisional integral for electron-ion collisional plasmas~\cite{Landau:1936dvu}.

First, we establish the differential cross section for MC-proton scattering~\cite{Davidson:2000hf},
\begin{equation}
    \frac{d\sigma}{dx}=\frac{\pi Q^2\alpha^2(1+x)}{2p^2(1-x)\left(1-x+\frac{k_D^2}{2p^2}\right)}.
\end{equation}
Here $p$ is the MC energy, $k_D$ is the inverse Debye screening length, and $x=\cos\Theta$ is the cosine of the MC deflection angle. In principle, $p$ and $x$ should be evaluated in the center-of-mass frame of the collision. However, to lowest order in $T/M$, where $T$ is the temperature of the dark plasma and $M$ is the mass of the nucleon, there is no energy exchange, and the center-of-mass frame coincides with the frame with the nucleon at rest. The slow energy exchange, induced by terms of order $T/M$, is anyway slower than the energy exchange induced by the coalescence processes, and so is negligible. Therefore, to this order, $p$ and $x$ can be interpreted as the relevant quantities in the laboratory frame.

We will work throughout in the approximation that $k_D\ll p$. While this may not necessarily be quantitatively accurate throughout the supernova core, the resulting collisional integral captures all the qualitative features of the trapping induced by Coulomb and self-interactions. In this approximation, the scattering is strongly forward, due to the denominator getting very small for $x\simeq 1$, so we can approximate
\begin{equation}
    \frac{d\sigma}{dx}=\frac{\pi Q^2\alpha^2}{p^2(1-x)\left(1-x+\frac{k_D^2}{2p^2}\right)}.
\end{equation}

In order to obtain the Fokker-Planck equation, we need to obtain the mean momentum and squared momentum exchanged between the MCP and the medium. In the approximation of elastic MCP scattering, after the collision, the MCP momentum is simply changed in direction by the angle $\Theta$. Therefore, the average loss in momentum in such a collision is $\langle -\delta\bp\rangle=\bp(1-x)$. The average momentum loss rate is then
\begin{equation}
    -\langle \frac{\delta \bp}{\delta t}\rangle n_p \int_{-1}^{+1} dx \frac{d\sigma}{dx} \bp (1-x)=\frac{\pi Q^2\alpha^2 n_p \Lambda}{p^2}\bp,
\end{equation}
where $n_p$ is the proton number density, and we have introduced the Coulomb logarithm~\cite{Landau:1936dvu},
\begin{equation}
    \Lambda=\int_{-1}^{+1}\frac{dx}{1-x+\frac{k_D^2}{2p^2}}\simeq\log\left(\frac{4p^2}{k_D^2}\right).
\end{equation}
Our scheme of approximation is logarithmically accurate and keeps only terms proportional to this logarithm. The mean squared change in the momentum is also needed; in a collision at a fixed deflection angle $\Theta$, the change in the transverse momentum is $|\delta\bp_{\bot}|^2=p^2 \sin^2\Theta\simeq 2p^2(1-\cos\Theta)$, so by the same averaging we find 
\begin{equation}
    \langle \frac{\delta p_i \delta p_j}{\delta t}\rangle =\pi  Q^2 \alpha^2 n_p \Lambda \left(\delta_{ij}-\frac{p_ip_j}{p^2}\right).
\end{equation}

The Coulomb collisional integral can now be written as
\begin{equation}
    \left(\frac{\partial f_\bp}{\partial t}\right)_\mathrm{Coul}=\frac{\partial}{\partial p_i}\left[-\langle \frac{\delta p_i}{\delta t}\rangle f_\bp+\frac{1}{2}\frac{\partial}{\partial p_j}\left(\langle\frac{\delta p_i \delta p_j}{\delta t}\rangle f_\bp\right)\right].
\end{equation}
This form of the collisional integral is general. However, we can provide an explicit expression for the azimuthally symmetric case of interest here, where $f_\bp=f(p,\bn)$. In this case, the derivatives can be expressed in polar coordinates. Terms arising from differentiating the Coulomb logarithm are not logarithmically enhanced and can therefore be discarded. Thus, we are finally led to the form
\begin{equation}
    \left(\frac{\partial f(p,\bn)}{\partial t}\right)_\mathrm{Coul}=\frac{\pi  Q^2 \alpha^2 n_p \Lambda}{2p^2 \sin\theta}\frac{\partial}{\partial \theta}\left[\sin\theta\frac{\partial f}{\partial\theta}\right],
\end{equation}
where $\theta$ is the polar angle of the particle momentum. This form of the collisional integral makes several features of the process explicit; first, the order of magnitude of the interaction rate is given by $ Q^2 \alpha^2 n_p \Lambda/T^2$, as expected from simpler arguments. Since we are interested only in logarithmic accuracy, we can safely replace in the Coulomb logarithm $\Lambda\simeq \log(4T^2/k_D^2)$. It is also evident from this form that the collisional integral conserves particle number and energy, while it tends to isotropize the angular distribution of the particles.

While this form of the collisional integral is general, in reality the MCPs are so strongly interacting that they behave as a fluid. Coulomb collisions act as a sink of momentum for the fluid; so we need to determine the transfer of momentum to the dark fluid
\begin{equation}
    \frac{d\boldsymbol{\mathcal{P}}}{dVdt}=\int \frac{d^3\bp}{(2\pi)^3}4\bp \frac{\pi  Q^2 \alpha^2 n_p \Lambda}{2p^2 \sin\theta}\frac{\partial}{\partial \theta}\left[\sin\theta\frac{\partial}{\partial\theta} \left(e^{-\frac{\gamma p (1-v\cos\theta)}{T}}\right)\right],
\end{equation}
where the factor $4$ accounts for the two spin states and the particle/antiparticle species. The integral is easily computed in the limit of small velocities, yielding
\begin{equation}
    \frac{d\boldsymbol{\mathcal{P}}}{dVdt}=-\frac{4 Q^2\alpha^2 n_p \Lambda T^2 \mathbf{v}}{3\pi}.
\end{equation}
This amounts essentially to a drag term that tends to slow down the fluid as it escapes the PNS core.

\section{D.~Comparison with previous bounds}

\begin{figure*}
    \includegraphics[width=\textwidth]{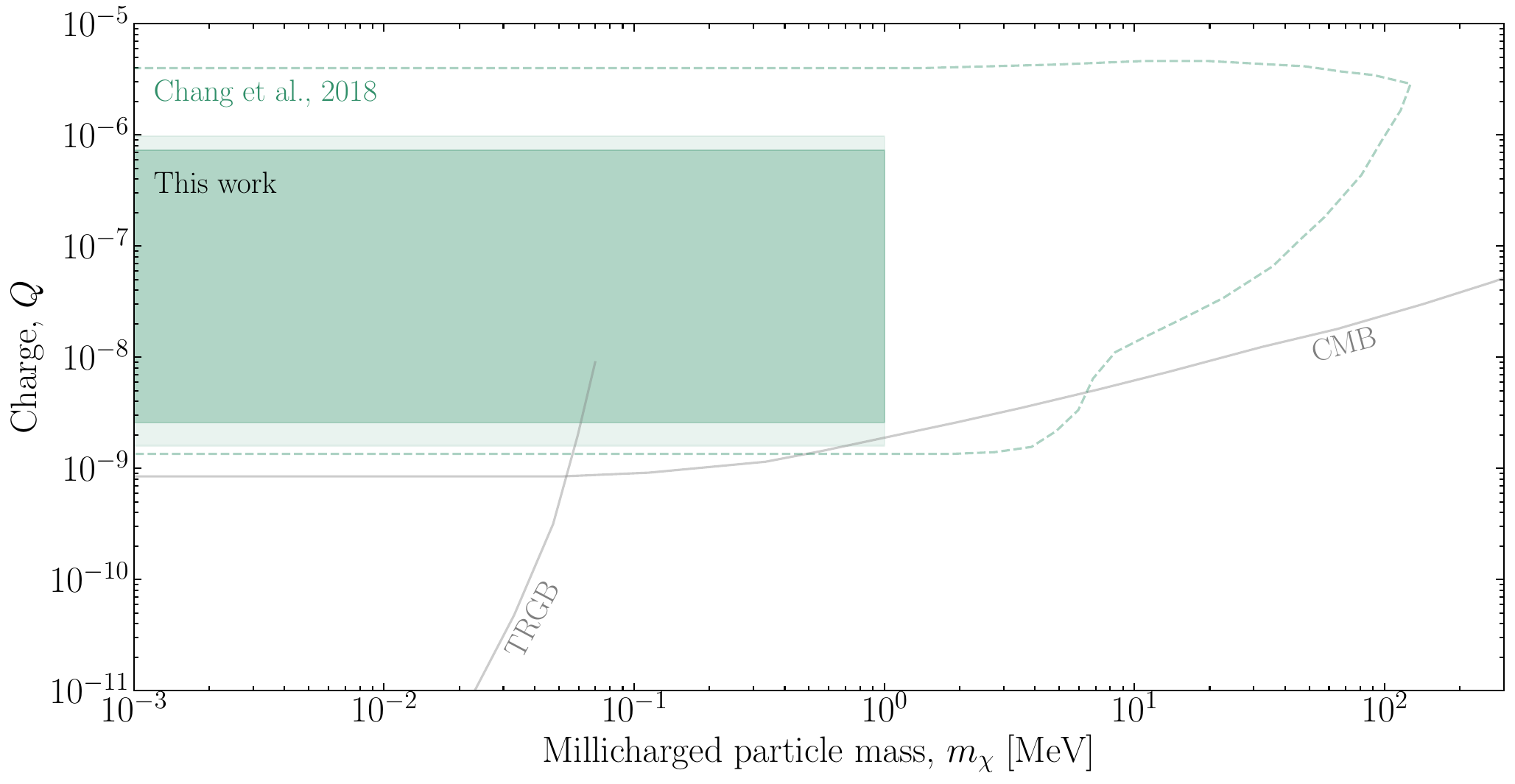}
    \caption{Comparison of the bounds obtained in this work with the latest SN bounds on MCPs~\cite{Chang:2018rso}, which were obtained neglecting self-interactions and any potential impact of SN magnetic fields. Our cooling bounds are shown both for cold (darker shading) and for the hot (lighter shading) model. We show in gray bounds from tip of the red-giant branch (TRGB)~\cite{Fung:2023euv} and from the Cosmic Microwave Background (CMB)~\cite{Adshead:2022ovo} corresponding to a change in the effective number of degrees of freedom $\Delta N_{\rm eff}=0.3$. CMB bounds assume $\alpha_\chi=7\times 10^{-4}$; the cooling bounds obtained in this work are independent of $\alpha_\chi$ provided that $\alpha_\chi\gtrsim 10^{-6}-10^{-5}$, guaranteeing that the MCPs escape as a fluid.}\label{fig:comparison_bounds}
\end{figure*}

The bounds we have obtained in this work are the first ones on self-interacting MCPs accounting for the hydrodynamical nature of their emission. Nevertheless, for practical purposes it may be worthwile comparing them to the previous SN bounds obtained neglecting MCP self-interactions. Notice that, if MCPs truly have no self-interactions, then magnetic fields might affect these bounds, an effect that has never been studied and that we leave for future work.

In Fig.~\ref{fig:comparison_bounds}, we show the cooling bounds obtained in this work (the energy deposition bounds, as shown in the main text, are comparable). We have assumed everywhere that the mass of the MCP $m_\chi$ is much smaller than the typical SN temperature in the emission region, so that it can be effectively considered massless for the emission rates. For the free-streaming regime, no further assumption is needed, since the cooling rate is independent of the further evolution of the fluid. Since the bulk of the emission happens below $10$~km, where the plasma frequency is about $10$~MeV, the approximation of massless MCP can be reasonably trusted for $m_\chi\lesssim 5$~MeV. For the trapping regime, one must require that $m_\chi\lesssim T_\chi$, the fluid temperature, out to the surface emission radius of the fluid; from Fig.~2 in the main text, we see that this corresponds to a fluid temperature $T_\chi\sim 5$~MeV. In addition, since the surface emission radius is closer to $20$~km, where the plasma frequency is of the order of $2-3$~MeV, the approximation of massless MCP can be trusted up to $m_\chi\sim 1$~MeV. Therefore, we extend our bounds up to $m_\chi=1$~MeV conservatively also for the free-streaming regime, since the shape of the bounds that are expected to extend up to hundreds of MeV requires a dedicated analysis; our work is the first step in this direction. %Yet higher masses could be excluded, but this requires a dedicated analysis in which the equation of state of a dark fluid with finite-mass particles is considered; our work is the first step in this direction.
For the bounds on non-self-interacting MCPs, taken from Ref.~\cite{Chang:2018rso}, we consider the most conservative of their numerical models in the trapping region, from Ref.~\cite{Fischer:2016cyd}. The so-called ``fiducial'' model, in which trapping seems to kick in at smaller couplings, is an analytical model (see Ref.~\cite{Chang:2016ntp}) which fails to fit precisely enough the behavior of the temperature and density.

Compared to the bounds on non-self-interacting MCPs, we find that the free streaming is more or less consistent, albeit our most conservative free-streaming bound is actually weaker than Ref.~\cite{Chang:2018rso} by a factor~2. This however is not connected with the hydrodynamical nature of the emission---the rate of energy emission in the free-streaming region is independent of the fluid dynamics---but rather with the SN profile used. On the other hand, in the trapping region the bounds differ by nearly as much as an order of magnitude, a difference mostly due to the neglect of self-annihilations in Ref.~\cite{Chang:2018rso}.

\bibliographystyle{bibi}
\bibliography{References}

\end{document}